\documentclass[reprint,aps]{revtex4-2}
\usepackage{amsmath}
\usepackage{amssymb}
\usepackage{nccmath}
\usepackage{bm}
\usepackage{float}
\usepackage{slashed}
\usepackage[export]{adjustbox}
\usepackage{comment}
\usepackage{graphicx}
\usepackage[colorlinks=true,linktocpage=true,
linkcolor=blue,citecolor=blue]{hyperref}

\begin{document}
		\large
	\title{\bf{Chirality dependence of thermoelectric response in a thermal QCD medium}}
	\author{Debarshi Dey\footnote{ddey@ph.iitr.ac.in, debs.mvm@gmail.com}~~and~~Binoy Krishna
		Patra\footnote{binoy@ph.iitr.ac.in}\vspace{0.1in}\\
		\textit{Indian Institute of Technology Roorkee, Roorkee 247667, India}}
	\date{}
	
	\begin{abstract}
	 The lifting of the degeneracy between $L$- and $R$-modes
of massless flavors in a weakly magnetized 
thermal QCD medium leads to a novel phenomenon of chirality dependence of 
the thermoelectric tensor, whose diagonal and non-diagonal elements are 
the Seebeck and Hall-type Nernst coefficient, respectively. 
Both coefficients in $L$-mode have 
been found to be greater than their counterparts in $R$-mode, however the disparity
is more pronounced in the Nernst coefficient.
Another noteworthy observation is the impact of the dimensionality of
temperature ($T$) profile on the Seebeck coefficient, wherein we find that
the coefficient magnitude is significantly enhanced ($\sim$ one order of magnitude) in the 2-D setup, compared to a 1-D $T$ profile. Further, the chiral dependent quasifermion masses constrain the range of magnetic field ($B$) and $T$ in a manner so as to enforce the weak magnetic field ($eB\ll T^2$) condition.
	\end{abstract}
	\maketitle
\emph{\bf Introduction-} Nuclear matter at a very high 
temperature and/or baryon density is conceived in terms 
of deconfined quarks and gluons (dubbed as QGP). Transport coefficients serve as input for modelling the flow of such matter created in relativistic 
heavy ion collision experiments\cite{Karsch:PLB3'2008,Sasaki:NPA832'2010,Finazzo:JHEP51'2015}.  
These experiments indicate that the created matter is very nearly an ideal fluid with the viscosity 
being close to the conjectured lower bound\cite{kovtun:PRL94'2005} arrived at from AdS/CFT correspondence. 
The QGP may be exposed to magnetic fields ($B$) arising from non-central nucleus-nucleus collisions\cite{sokov:IJMPA24'2009,tuchin:PRC82'2010}; its strength depending on the time scale of evolution.  
A strong $B$ provides the ground for probing the topological properties of QCD 
vacuum\cite{kharzeev:NPA803'2008,fukushima:PRL120'2018}, whereas weak $B$ yields some novel 
phenomenological consequences through the lifting of degeneracy between left and right handed quarks. Our aim 
is to explore the consequence of this splitting on the thermoelectric 
response of the medium.

In weak $B$ regime ($m_{0}^2\ll eB\ll T^2$), the thermoelectric 
response of the thermal QCD medium assumes a matrix structure :
\begin{equation}\label{matrix}
\begin{pmatrix}
E_x\\
E_y
\end{pmatrix}=\begin{pmatrix}
S & N|\bm B|\\
N|\bm B| & S
\end{pmatrix}\begin{pmatrix}
{\nabla T|}_x\\
{\nabla T|}_y
\end{pmatrix}.
\end{equation}
 The diagonal and nondiagnal elements are Seebeck ($S$)
 and Nernst ($N|\bm B|$) coefficients, respectively, which are
 the measures of `longitudinal' and `transverse' (analogous to Hall effect) induced electric fields. Large fluctuations in the initial energy density in the 
heavy-ion collisions\cite{Schenke:PRL108'2012} translate to significant temperature gradients between 
the central and peripheral regions of the produced fireball, providing the ideal ground to study thermoelectric phenomena.
 
 We also look at the impact of dimensionality of temperature profile (1-D/2-D) on the Seebeck coefficient.
  At strong $B$, fermions are constrained to move only in one
 dimension, {\em i.e.} only the lowest Landau level (LLL) is populated.
 When the strength of $B$ 
 decreases, higher Landau levels start getting occupied, consequently, the constraint 
 on motion of fermions is relaxed and the study of thermoelectric response with a 1-D or 2-D temperature profile becomes plausible. However, the Nernst effect is only manifested in 
 the weak $B$ regime, since the transverse current vanishes at strong $B$.\\

\noindent\emph{\bf Dispersion Relations in weak $B$: Chiral modes}- The dispersion relation of quarks is obtained from the zeros of the inverse
resummed quark propagator:
\begin{equation}
S^{-1} (K)= S^{-1}_0 (K) - \Sigma (K),
\end{equation}
where the  quark self-energy, $\Sigma (K)$ is to be calculated up to
one-loop from thermal QCD in weak $B$ and the bare quark propagator, 
$S_0(K)$ in weak $B$, up to power $|q_fB|$, is given by
\begin{equation}\label{fermion_prop_magnetic}
iS_{0}(K) = \frac{i\left(\slashed{K}+m_f\right)}{K^2-m_f^2} - \frac{\gamma_1\gamma_2\left(\gamma.K_{||}+m_f\right)}{\left(K^2-m_f^2\right)^2}(q_fB),
\end{equation}
which can be written in terms of the fluid four-velocity $u^{\mu}= (1,0,0,0)$ and $b^{\mu}=\frac{1}{B}\epsilon^{\mu\nu\rho\lambda}u^{\nu}F^{\rho\lambda}=(0,0,0,1)$, 
as,
\begin{equation}\label{prop_2}
iS_0(K) = \frac{i\left(\slashed{K}\right)}{K^2-m_f^2} - \frac{i\gamma_5[(K.b)\slashed{u}-(K.u)\slashed{b}]}{(K^2-m_f^2)^2}\big(q_f B\big).
\end{equation} 
The one loop quark self energy is then given by
\begin{align}\label{self_energy}
\Sigma(P) = & g^2 C_F T\sum_n\int\frac{d^3k}{(2\pi)^3}\gamma_{\mu}\Bigg(\frac{\slashed{K}}{(K^2-m_f^2)} -\nonumber\\
& \frac{\gamma_5[(K.b)\slashed{u}-(K.u)\slashed{b}]}{(K^2-m_f^2)^2}(|q_fB|)\Bigg)\gamma^{\mu}\frac{1}{(P-K)^2}
\end{align}
In a covariant tensor basis, the above self energy can be expressed in terms of structure constants 
$\mathcal{A}, \mathcal{B}, \mathcal{C}, \mathcal{D}$ as
\begin{equation}\label{52}
\Sigma(P) = -\mathcal{A}\slashed{P}-\mathcal{B}\slashed{u}-\mathcal{C}\gamma_5\slashed{u}-\mathcal{D}\gamma_5\slashed{b},
\end{equation} 
The self energy and the full propagator can then be rewritten in terms of 
projection operators $P_L =(\mathbb{I}-\gamma_5)/2$ and $P_R =(\mathbb{I}+\gamma_5)/2$ as
\begin{equation}\label{63}
\Sigma(P) = -P_R\slashed{A'}P_L - P_L\slashed{B'}P_R,
\end{equation}
\begin{equation}
S^{-1}(P) = P_R\slashed{L}P_L + P_L\slashed{R}P_R,
\end{equation}
\begin{equation}\label{prop_eff}
S(P) = \frac{1}{2}\Big[ P_L\frac{\slashed{L}}{L^2/2}P_R+\frac{1}{2}P_R\frac{\slashed{R}}{R^2/2}P_L\Big],
\end{equation}
where, 
\begin{align}
 L^2 &= (1+\mathcal{A})^2P^2 + 2(1+\mathcal{A})(\mathcal{B}+\mathcal{C})p_0\\\nonumber&-2\mathcal{D}(1+a)p_z+ (\mathcal{B}+\mathcal{C})^2-\mathcal{D}^2,
 \end{align}
 \begin{align}
  R^2 &= (1+\mathcal{A})^2P^2 + 2(1+\mathcal{A})(\mathcal{B}-\mathcal{C})p_0\\\nonumber&+2\mathcal{D}(1+\mathcal{A})p_z+ (\mathcal{B}-\mathcal{C})^2-\mathcal{D}^2.
 \end{align}
  
The $p_0=0$, ${\bf{p}}\rightarrow 0$ limit of the denominator of the effective propagator yields 
the quasiparticle masses as\cite{das:PRD97'2018,pushpa}
\begin{align}\label{mass}
&m_L^2=\frac{L^2}{2}\arrowvert_{p_0= 0, {|\bf{p}|}\rightarrow 0} = m_{th}^2 + 4g^2 C_F M^2,\\
&m_R^2=\frac{R^2}{2}|_{p_0= 0, {|\bf{p}|}\rightarrow 0} = m_{th}^2 - 4g^2 C_F M^2,
\end{align}
thus lifting the degeneracy. Here,
\begin{align}
 M^2 &= \frac{|q_fB|}{16\pi^2}\left(\frac{\pi T}{2m_f} -\text{ln} 2+ \frac{7\mu^2\zeta(3)}{8\pi^2T^2}\right),\\
m_{th}^2&=\frac{1}{8}g^2C_F\left(T^2+\frac{\mu^2}{\pi^2}\right).
\end{align}
The coupling constant, $g$ is used as in\cite{Ayala:PRD98'2018}. 
In the ultra relativistic limit, the chirality of a particle is the same as it's helicity, so that the 
right and left chiral modes can be thought of as the up/down spin projections in the direction of momentum ($s_z=\pm 1/2$) 
of the particle, suggesting that the medium generated mass of a collective fermion excitation at very high temperatures ($T\gg m_0$) is $s_z$  
(helicity)-dependent. Such $s_z$ dependent quasiparticle masses are already known in condensed matter 
systems\cite{spaek:PRL64'1990,Riseborough:PhilMag'2006}.\\

\noindent \emph{\bf The thermoelectric coefficients}- We make use of the Boltzmann transport equation to 
calculate the infinitesimal deviation from equilibrium of the system caused by the temperature gradient. 
\begin{equation}
p^{\mu}\frac{\partial f_i(x,p)}{\partial 
	x^{\mu}}+ q_iF^{\mu\nu}p_{\nu}\frac{\partial f_i(x,p)}{\partial p^{\mu}}=\left(\frac{\partial f_i}{\partial t}\right)_{\text{coll}}
\label{BTE}
\end{equation}
The highly non linear collision integral on the R.H.S. can be linearized using the relaxation time 
approximation which reads (suppressing the flavor index $i$)
\begin{equation}
\left(\frac{\partial f}{\partial t}\right)_{\text{coll}}\simeq -\frac{p^{\mu}u_{\mu}}{\tau}\delta f=-\frac{f-f_0}{\tau},
\end{equation}
 where, $\tau$ is the relaxation time\cite{Hosoya:NPB250'1985} and $f_0$ is the Fermi-Dirac distribution 
 function. $\delta f$ is then used to calculate the induced current which is set to zero (enforcing 
 the equilibrium condition) to evaluate the relevant response functions\cite{Callen} 
 (Seebeck and Nernst coefficients). A non-zero $\mu$ leads to a non-zero net thermocurrent and induces 
an electric field which grows until the thermocurrent is neutralised. In condensed matter systems, 
an electric field is applied externally to achieve the condition of zero thermocurrent and the coefficient
 (Seebeck/Nernst) is read off therefrom.

We use the following Ansatz\cite{Feng:PRD96'2017} to solve for $\delta f$ from Eq.\eqref{BTE}
\begin{equation}
f^{L/R}=f_0^{L/R}-\tau q\bm{E}\cdot\frac{\partial f_0^{L/R}}{\partial \bm{p}}-\bm{\chi}.\frac{\partial f_0^{L/R}}{\partial \bm{p}}\label{ansatz},
\end{equation}
where, the effect of $\mathbf{B}$ is encoded in $\bm{\chi}$ ($L/R$ denotes the handedness). We begin with a single flavor system. The components of $\bm{\chi}$, 
after some algebra, can be expressed conveniently in a matrix form 
\begin{multline*}
\begin{bmatrix}
\chi_x\\
\chi_y
\end{bmatrix}=\begin{bmatrix}
\frac{-\omega_c\tau^2}{1+\omega_c^2\tau^2}q & \frac{\omega_c\tau^2}{1+\omega_c^2\tau^2}q\\
\frac{-\omega_c\tau^2}{1+\omega_c^2\tau^2}q & -\frac{\omega_c^2\tau^3}{1+\omega_c^2\tau^2}q
\end{bmatrix}\begin{bmatrix}
E_x\\
E_y
\end{bmatrix}+\\\begin{bmatrix}
-\frac{\tau}{1+\omega_c^2\tau^2}\left(\frac{\epsilon-\mu}{T}\right) & -\left(\frac{\epsilon-\mu}{T}\right)\frac{\omega_c\tau^2}{1+\omega_c^2\tau^2}\\
\frac{\omega_c\tau^2}{1+\omega_c^2\tau^2}\left(\frac{\epsilon-\mu}{T}\right) & -\frac{\tau}{1+\omega_c^2\tau^2}\left(\frac{\epsilon-\mu}{T}\right)
\end{bmatrix}\begin{bmatrix}
\frac{\partial T}{\partial x}\\[0.3em]
\frac{\partial T}{\partial y}
\end{bmatrix}.
\end{multline*}
The induced 4-current is given by (We drop the $L/R$ notation for brevity):
\begin{equation}
J^{\mu}=qg\int \frac{d^3\mbox{p}}{(2\pi)^3\epsilon}p^{\mu}\left[\delta f-\overline{\delta f}\right],\label{current_def}
\end{equation}
where, $\overline{\delta f}$ denotes the contribution from antiparticles. For $\bm{\nabla T}=\frac{\partial T}{\partial x}\bm{\hat{x}}+\frac{\partial T}{\partial y}\bm{\hat{y}}$ the equilibrium condition is $J_x=J_y=0$, which yields the following equations
\begin{align*}
&\left[(C_1)E_x+(C_2)E_y+(C_3)\frac{\partial T}{\partial x}+(C_4)\frac{\partial T}{\partial y}\right]=0\\
&\left[-(C_2)E_x+(C_1)E_y-(C_4)\frac{\partial T}{\partial x}+(C_3)\frac{\partial T}{\partial y}\right]=0
\end{align*}
Solving for $\bm{E}$ in terms of $\bm{\nabla}T$ leads to the structure
\begin{equation}
\begin{pmatrix}
E_x\\
E_y
\end{pmatrix}=\begin{pmatrix}
S & N|\bm{B}|\\
-N|\bm{B}| & S
\end{pmatrix}\begin{pmatrix}
\frac{\partial T}{\partial x}\\[0.3em]
\frac{\partial T}{\partial y}
\end{pmatrix}.
\end{equation}
 with
\begin{align}
S&=-\frac{C_1C_3+C_2C_4}{C_1^2+C_2^2},\label{60}\\[0.3em]
N|\bm{B}|&=\frac{C_2C_3-C_1C_4}{C_1^2+C_2^2}.\label{61}
\end{align}
where,
\begin{align*}
C_1=q&\int \mbox{dp}\,p^4 \frac{\tau}{\epsilon^2(1+\omega_c^2\tau^2)}\big\{f_0(1-f_0)\\
+&\bar{f_0}(1-\bar{f_0})\big\},\\[0.3em]
C_2=q&\int \mbox{dp}\,p^4 \frac{\omega_c\tau^2}{\epsilon^2(1+\omega_c^2\tau^2)}\big\{f_0(1-f_0)\\
-&\bar{f_0}(1-\bar{f_0})\big\},\\[0.3em]
C_3=\beta& \int \mbox{dp}\,p^4\frac{\tau}{\epsilon^2(1+\omega_c^2\tau^2)}\big\{(\epsilon+\mu)\bar{f_0}\\&(1-\bar{f_0})
-(\epsilon-\mu)f_0(1-f_0)\big\},\\[0.3em]
C_4=\beta&\int \mbox{dp}\,p^4\frac{\omega_c\tau^2}{\epsilon^2(1+\omega_c^2\tau^2)}\big\{-(\epsilon+\mu)\bar{f_0}\\
&(1-\bar{f_0})-(\epsilon-\mu)f_0(1-f_0)\big\}.
\end{align*}
For the physical medium consisting of $u$ and $d$ quarks, the total currents are given as
\begin{align}
J_x=&\sum_{a=u,d}\left[(I_1)_aE_x+(I_2)_aE_y+(I_3)_a\frac{\partial T}{\partial x}+(I_4)_a\frac{\partial T}{\partial y}\right]\\
J_y=&\sum_{a=u,d}\left[-(I_2)_aE_x+(I_1)_aE_y- (I_4)_a\frac{\partial T}{\partial x}+(I_3)_a\frac{\partial T}{\partial y}\right].
\end{align}
This ultimately leads to 
\begin{align}
S&=-\frac{K_1K_3+K_2K_4}{K_1^2+K_2^2}\\
N|\bm{B}|&=\frac{K_2K_3-K_1K_4}{K_1^2+K_2^2},
\end{align}
where, 
\begin{align}
K_1&=\sum_{a=u,d}(I_1)_a\,,\qquad K_2=\sum_{a=u,d}(I_2)_a\,,\nonumber\\[0.3em]
K_3&=\sum_{a=u,d}(I_3)_a\,,\qquad \, K_4=\sum_{a=u,d}(I_4)_a.\nonumber
\end{align}
As mentioned earlier, the Seebeck coefficient can also be evaluated with a 1D temperature profile\cite{Bhatt:PRD99'2019,Dey:PRD102'2020},
{\em e.g.} $\bm{\nabla}T=\frac{\partial T}{\partial x}\bm{\hat{x}}$. For the composite medium, it is
given as
\begin{equation}\label{sc1d}
S_{1D}=\frac{1}{T}\frac{\sum\limits_{a=u,d}q_a(I_2)_a}{\sum\limits_{a=u,d}q_a^2\,(I_1)_a}=\frac{\sum\limits_{a} S_a\,q_a^2(I_1)_a}{\sum\limits_
	{a}q_a^2(I_1)_a},
\end{equation}
with
\begin{align*}
I_1=q&\int \mbox{dp}\,p^4 \frac{\tau}{\epsilon^2(1+\omega_c^2\tau^2)}\big\{f_0(1-f_0)\\
+&\bar{f_0}(1-\bar{f_0})\big\},\\[0.3em]
I_2=q&\int \mbox{dp}\,p^4 \frac{\tau}{\epsilon^2(1+\omega_c^2\tau^2)}\big\{(\epsilon-\mu)f_0(1-f_0)\\
-&(\epsilon+\mu)\bar{f_0}(1-\bar{f_0})\big\}
\end{align*}
Thus, the total Seebeck coefficient is expressed as a weighted average of the single component coefficients.
Such a mathematical structure is absent for the 2D $T$ profile. Since the Nernst coefficient relates the 
induced electric field and the temperature gradient in mutually transverse directions, it emerges along
with the Seebeck coefficient naturally in the 2D setup.
\vspace{5mm}
\begin{figure}[H]
	\centering
	\includegraphics[width=8cm]{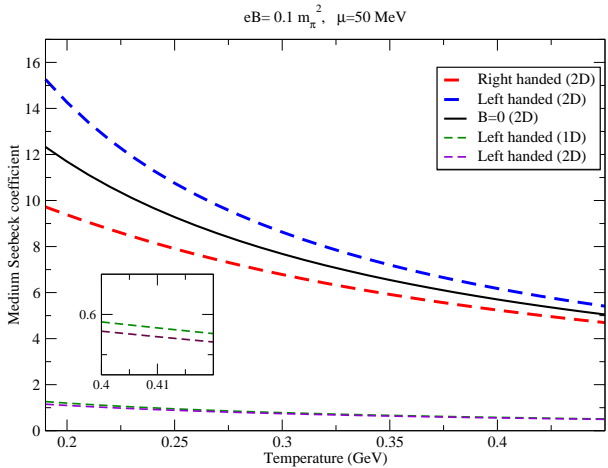}
	\vspace{-0.2cm}
	\caption{$L$ and $R$ mode medium Seebeck 
		coefficients  as a function of $T$.
	}
	\label{seebeck_med}
\end{figure}

\begin{figure}[H]
	\centering
	\includegraphics[width=8cm]{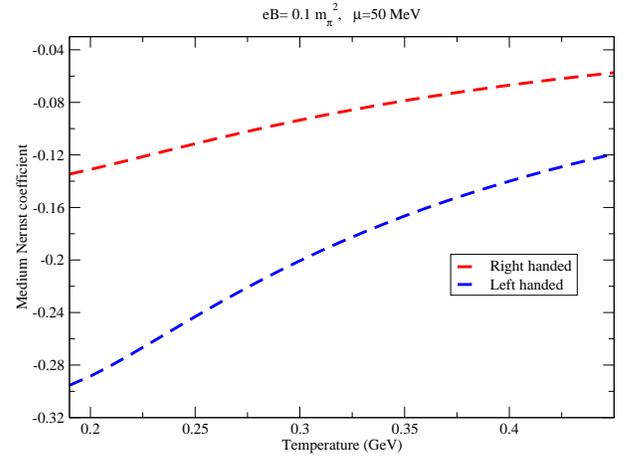}
	\vspace{-0.2cm}
	\caption{$L$ and $R$ mode medium Nernst
		coefficients  as a function of $T$.
	}
	\label{nernst_med}
\end{figure}
Figures \eqref{seebeck_med} and \eqref{nernst_med} show the variation of Seebeck and Nernst coefficients 
of the medium with temperature. It can be seen that the magnitude of the induced electric field in the 
longitudinal (along $\bm{\nabla T}$) and transverse (perpendicular to $\bm{\nabla T}$) directions shows 
similar trends as far as variation with temperature is considered; for both the modes, the magnitudes 
decrease with temperature. Also, for both the coefficients, the $L$ mode elicits a larger comparative 
response. This can be understood from a numerical perspective. Each of the integrals $C_1$, $C_2$, $C_3$, $C_4$ 
are decreasing functions of the effective mass. However, the extent of decrease follows the hierarchy 
$\Delta C_2>\Delta C_1>\Delta C_4>\Delta C_3$, where $\Delta C$ denotes change in the value of the integral 
due to a given change in mass. The mathematical expressions of $S$ and $N|\bm{B}|$ then imply that whereas 
both the numerator and denominator of the expressions decrease with increasing mass, the denominator 
decreases by a larger amount (because of the presence of $C_1$ and $C_2$) than the numerator. The value 
of the fraction, therefore, increases with increasing mass. For comparison, the $B=0$ case is also shown 
for the Seebeck coefficient. The Nernst coefficient is, however zero for $B=0$, as should be the case. 

Fig.\eqref{seebeck_med} shows the impact of the dimension of $T$ profile on the magnitude and temperature 
behaviour of the Seebeck coefficient. For the 1D setup, Both the $I_1$ and $I_2$ integrals are decreasing 
functions of mass, \emph{i.e.} their values increase in going to the $R$ mode from the $L$ mode. The 
increase is however greater for $I_1$, compared to $I_2$ and hence, Eq.\eqref{sc1d} dictates the hierarchy 
that is observed in Fig.\eqref{seebeck_med}. As can be seen, there is almost an order magnitude increase 
in the medium Seebeck coefficient values in the 2-D case. The hierarchy with respect to magnitudes 
remains the same in the 1-D case with the $L$ mode lying above the $R$ mode for the entire temperature 
range considered. Compared to the 2-D case, there is stark contrast regarding the extent of splitting 
(in the Seebeck coefficient magnitude) witnessed in the 1-D setup with a maximum relative difference 
of 24.5\% between the $L$ and $R$ modes (57.1\% in the 2-D case). Thus, the magnitude as well as sensitivities 
 to mass and temperature of the thermoelectric response are heightened in the 2-D temperature profile. 

\begin{figure}[H]
	\includegraphics[width=9cm,height=7.5cm,inner]{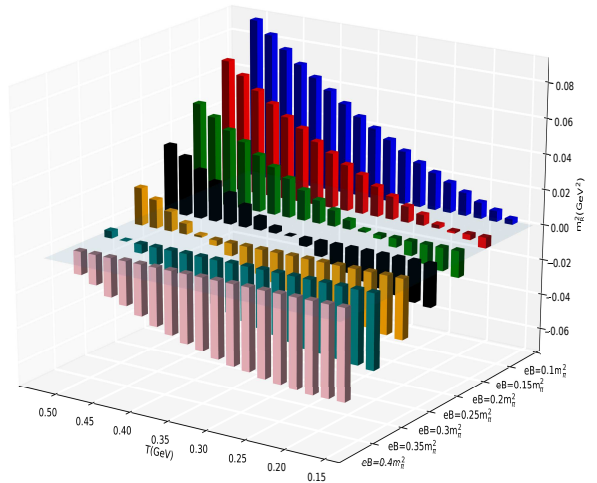}
	\vspace{-0.5cm}
	\caption{Medium generated mass (squared) of $R$ mode $u$ quark as a function of $T$ and $B$.}
	\label{3d}
\end{figure}
We found that the mass (squared) of the $R$ mode fermion, (Eq.\eqref{mass}) comes out to be negative 
for a certain range of combinations of $T$ and $B$ values, which is brought out by Fig.\eqref{3d}. As 
the magnetic field is increased, the temperature (above $T_c\sim 155$ MeV) upto which $m_R^2$ is negative, 
also increases. This suggests that the perturbative framework used by us to study the chirality dependence 
of the thermoelectric response is valid only at regions sufficiently far ($\sim$200 MeV) from the crossover 
region in the QCD phase diagram for $|eB|>0.2 m_{\pi}^2$. Another way to look at it is that the condition 
$eB\ll T^2$ is strictly enforced. For $T>T_c$, we find from Fig.\eqref{3d} that 
$\left|\frac{eB}{T^2}\right|_{\text{max}}\sim 0.07$ for $eB=0.1m_{\pi}^2$ and $\sim 0.03$ for $eB=0.35m_{\pi}^2$. 
For values of $T$ and $B$ leading to higher values of $\left|\frac{eB}{T^2}\right|$, $m_R^2$ is negative and thus unphysical.


 The Seebeck coefficient does not have an explicit $B$ dependence; its $B$ dependence stems from that of 
 the quasiparticle mass. The Nernst coefficient additionally carries an explicit $B$ dependence. As such, 
 its sensitivity to changes in magnetic field strength is comparatively more pronounced. The maximum 
 percentage difference in magnitudes between the $L$ and $R$ modes for the Seebeck and Nernst coefficients 
 is 57.1\% and 118.6\%, respectively.

\bibliography{mybib}

\end{document}